\documentclass[aps,prl,twocolumn,groupedaddress]{revtex4-1}

\usepackage{graphicx}
\usepackage{morefloats}
\usepackage{color}
\usepackage{amssymb}
\usepackage{amsmath}
\usepackage{float}
\usepackage{bm}
\usepackage{amsmath}
\usepackage{epstopdf}

\begin{document}

\newcommand{\uvec}[1]{\hat{\mathbf{#1}}}

\newcommand{\ham}{\mathcal{H}} 
\newcommand{\Q}{\mathcal{Q}} 

\title{Planar Hall driven torque in a FM/NM/FM system}

\author{Christopher Safranski}
\affiliation{IBM T. J. Watson Research Center, Yorktown Heights, New York 10598, USA}
\author{Jun-Wen Xu}
\author{Andrew D. Kent}
\affiliation{Center for Quantum Phenomena, Department of Physics,New York University, New York, NY 10003, USA}
\author{Jonathan Z. Sun}
\affiliation{IBM T. J. Watson Research Center, Yorktown Heights, New York 10598, USA}


\begin{abstract}

An important goal of spintronics is to covert a charge current into a spin current with a controlled spin polarization that can exert torques on an adjacent magnetic layer. Here we demonstrate such torques in a two ferromagnet system. A CoNi multilayer is used as a spin current source in a sample with structure CoNi/Au/CoFeB. Spin torque ferromagnetic resonance is used to measure the torque on the CoFeB layer. The response as a function of the applied field angle and current is consistent with the symmetry expected for a torques produced by the planar Hall effect  originating in CoNi. We find the strength of this effect to be comparable to that of the spin Hall effect in platinum, indicating that the planar Hall effect holds potential as a spin current source with a controllable polarization direction.

\end{abstract}

\flushbottom
\maketitle



The manipulation of magnetization through the use of spin torques \cite{Slonczewski1996} and the conversion of charge to spin current  are intensely studied areas of condensed matter physics. Investigation of these topics may pave the way towards energy efficient nanodevices    for memory \cite{Ikeda2010,Worledge2011} and novel computing applications \cite{Dieny2019,Torrejon2017,Camsari2018}. Spin orbit torques produced by the spin Hall  and Rashba effects in non-magnetic (NM) materials have been explored for switching of ferromagnets \cite{Liu2012,Miron2011} (FM) and spin torque oscillators \cite{Demidov2012,Duan2014}. The polarization of the spin currents in these systems has been dictated by sample geometry and crystal structure, limiting the form of torques that can be produced. Expansion of the class of materials and available spin polarization  has been investigated through the means of manipulating crystal symmetry \cite{MacNeill2016} and spin rotation \cite{Humphries2017,Baek2018} at interfaces. Additionally, spin orbit torques produced in FM materials \cite{Kurebayashi2014,Haidar2019,Wang2019,Taniguchi2015} have also gained interest.


In FM materials, it has  been proposed that  the anomalous (AHE) and planar (PHE) Hall effects can be used to produce spin current with a controllable spin polarization direction \cite{Taniguchi2015}. While the AHE has been experimentally  shown to inject spin current from one FM to another \cite{Gibbons2018,iihama2018spin,Seki}, torques from the PHE have not been observed in the two FM system. In this letter, we demonstrate that the PHE can be used to  inject spin current from one ferromagnet into another in a FM/NM/FM system.    In previous studies of a single FM/NM system, the planar Hall effect has been  shown to produce spin polarization with out of plane components and a unique angular symmetry \cite{Safranski2018}.   In the two FM system, we find this angular symmetry can be preserved and the torque strength is comparable  to spin Hall materials. Combined with the  controllable nature of the spin current  polarization, spin torques from PHE can be used to  broaden the  material systems and available symmetries for spintronics research.

\begin{figure*}[tbp]
\includegraphics[width=\textwidth]{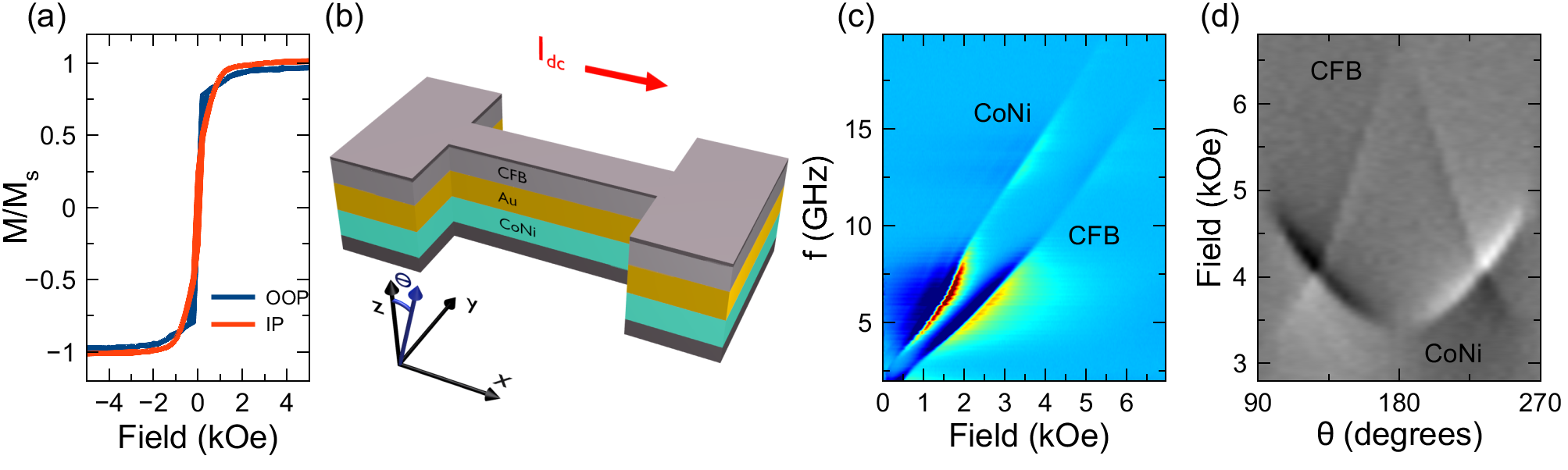}
\caption{(a) Vibrating sample magnetometry data for the film stack used showing saturation along the  axes. (b) Sample geometry and coordinate system. (c) ST-FMR voltage from self-rectification shown as a contour plot against frequency and field with $\theta$=205 degrees. $\theta$= 0 deg corresponds to sample-normal. (d) ST-FMR voltage as a function of $\theta$ and applied field, with a 14\,GHz  applied microwave current.   }
\label{fig1}
\end{figure*}

The planar Hall effect produces a charge current $\mathbf{J}_{\mathrm{PHE}}=\Delta \sigma_{\mathrm{AMR}}(\uvec{m} \cdot \mathbf{E}) \uvec{m}$ flowing in the FM parallel to its magnetization \cite{Taniguchi2015,McGuire1975}, where $\mathbf{E}\approx E\uvec{x}$ is the applied electric field and $\Delta\sigma_{\mathrm{AMR}}$ is anisotropic part of the FM conductivity. The charge current creates a spin current with  a spin polarization aligned with the magnetization \cite{Kokado2012,Taniguchi2015}.  From theoretical calculations \cite{Amin2018,Belashchenko2018},  the charge to spin conversion efficiency in materials such as Co is expected to be on the order of the  spin Hall effect in Pt.

To study PHE driven torques, we choose a  CoNi superlattice since it has a relatively strong PHE \cite{McGuire1975} and can be grown with perpendicular anisotropy \cite{Mangin2006,Arora2017}. Spin current is injected from this layer into an adjacent CoFeB (CFB) layer with relatively weak PHE \cite{Seemann2011}(See Supplementary Note 1).  Using magnetron sputtering, we deposit  Ta(3)/Pt(3)/CoNi/Au(3)/CFB(1.5)/Ta(3) (in nm), where the CoNi superlattice is [Co(0.65)/Ni(0.98)]$_{2}$/Co(0.65), onto an oxidized silicon substrate.   FM layer thicknesses are chosen such that CoNi is perpendicularly magnetized with the additional anisotroy generated by the Pt \cite{Carcia1988} layer, while the CFB is in-plane. The Au layer is used to break the exchange coupling between FM layers while allowing spin current carried by conduction electrons to pass. Its thickness is chosen such that the RKKY coupling across Au is weak \cite{Parkin1994}. Vibrating sample magnetometry measurements  in  Fig.\,\ref{fig1}(a) show that a 2\,kOe applied field can saturate both layers' magnetization along the applied field  for  in-plane and out-of-plane directions.  

Using E-beam lithography and ion milling techniques, we pattern 400\,nm wide 3\,$\mu$m long bridges into the film and encapsulate it with 40\,nm of SiN in situ. The resulting structure is schematically represented in Fig.\,\ref{fig1}(b). In this design, the leads are patterned from the same material as the bridges. The width of the leads is chosen such that the current density is too low to produce parasitic signals. 

In order to detect spin current injection, we employ spin torque ferromagnetic resonance (ST-FMR) techniques \cite{Tulapurkar2005,Sankey2006}. An amplitude modulated microwave current is applied directly to the device with a  modulation frequency of 1117\,Hz. Rectified voltages produced from ferromagnetic resonance are then measured using lock-in techniques. To produce additional spin currents, a DC current is supplied to the sample through a bias tee. Any resulting damping-like torques will then modify the FMR resonance linewidth \cite{Petit2007,iihama2018spin,Safranski2018}. 

Figure\,\ref{fig1}(c)  shows a contour plot of the measured ST-FMR signal as a function of microwave frequency and applied field at $\theta$=205 degrees. We observe two distinct resonances corresponding to the two ferromagnetic layers. In order to determine which layer each branch is associated with, we measure the resonance field as a function of applied field angle $\theta$ in the $xz$ plane shown as a contour plot in Fig.\,\ref{fig1}(d). A strong angular dependence is observed when rotating from in plane to out of plane. Since the CFB was chosen to be in-plane magnetized and CoNi to be out-of-plane, the angular dependence allows  the identification of the peaks shown in Figs.\,\ref{fig1}(c,d).  

We then determine the damping-like torques by a measurement of the resonance linewidth as a function of DC current.  ST-FMR measurements are performed at 14\,GHz to be in a field range where both FM are nearly collinear with the applied magnetic field. In this configuration, the spin current polarization from CoNi will be nearly collinear with the CFB layer magnetization. The resulting effect on the FMR resonance linewidth is maximal for this relative orientation between the spin current and  CFB magnetization direction.  To the leading order, the overall angular dependence for planar Hall driven torques with CFB and CoNi moments in the $xz$ plane will then be determined by the spin current  projection on the CFB interface, resulting in a dependence proportional to  $(\uvec{m} \cdot \uvec{x}) (\uvec{m} \cdot \uvec{z})=\cos(\theta)\sin(\theta)$ \cite{Safranski2018}.

\begin{figure*}[tbp]
\includegraphics[width=\textwidth]{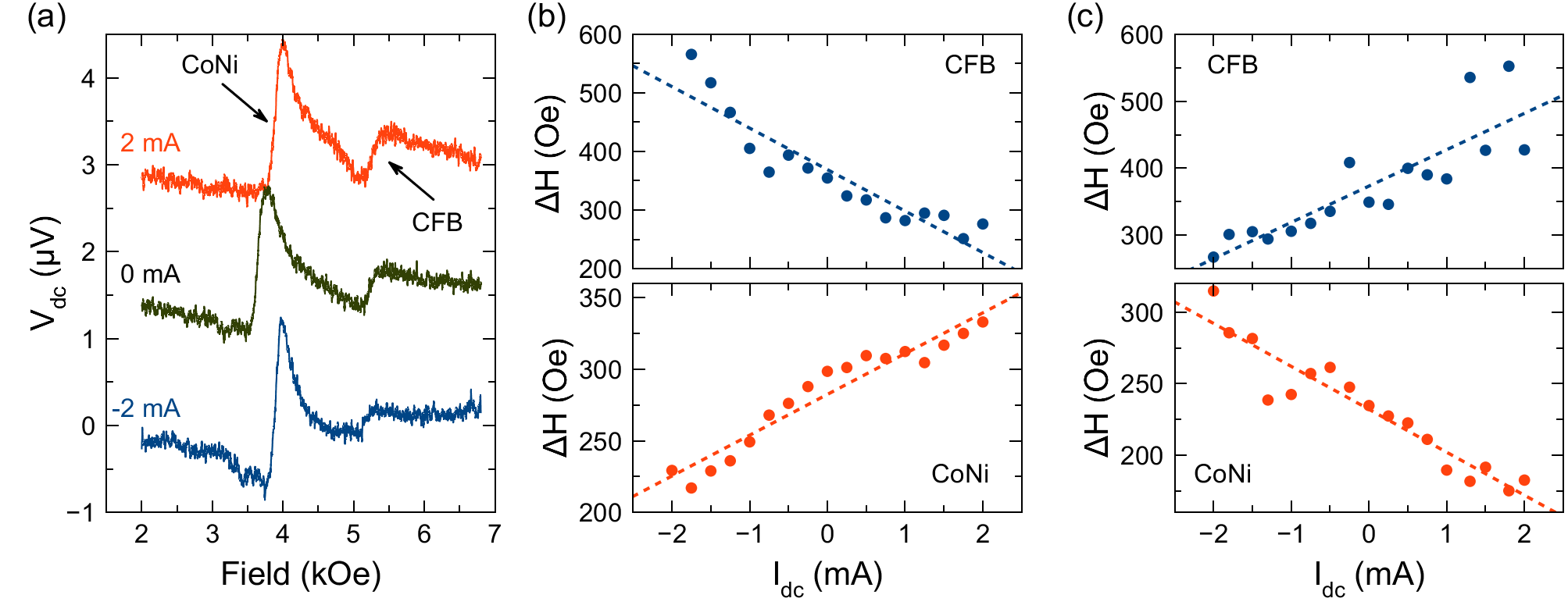}
\caption{ 
(a) ST-FMR signal at $\theta$=330 degrees for three different DC bias.  Resonance linewidth for each layer as a function of DC bias at (b) $\theta$=330 degrees and (c) $\theta$=205 degrees. 
}
\label{fig2}
\end{figure*}

Figure\,\ref{fig2}(a) shows three ST-FMR traces taken at different DC bias at $\theta$=330 degrees. We observe that the linewidth of both resonance  peaks is modified by the application of current. Fitting these to the sum of Lorentzian and anti-Lorentzian functions, we see a linear change in linewidth in Fig.\,\ref{fig2}(b) for both resonances. Further, we observe that when the CFB layer's linewidth narrows, the linewidth of the CoNi layer increases. From the angular dependence of the planar Hall current, if the magnetization is rotated across the $z$ axis to $\theta$=205 degrees, we would expect to see a change in the sign of the linewidth vs bias slope. Figure\,\ref{fig2}(c) shows that the slope does indeed change sign. 

\begin{figure}[pt]
\includegraphics[width=0.5\textwidth]{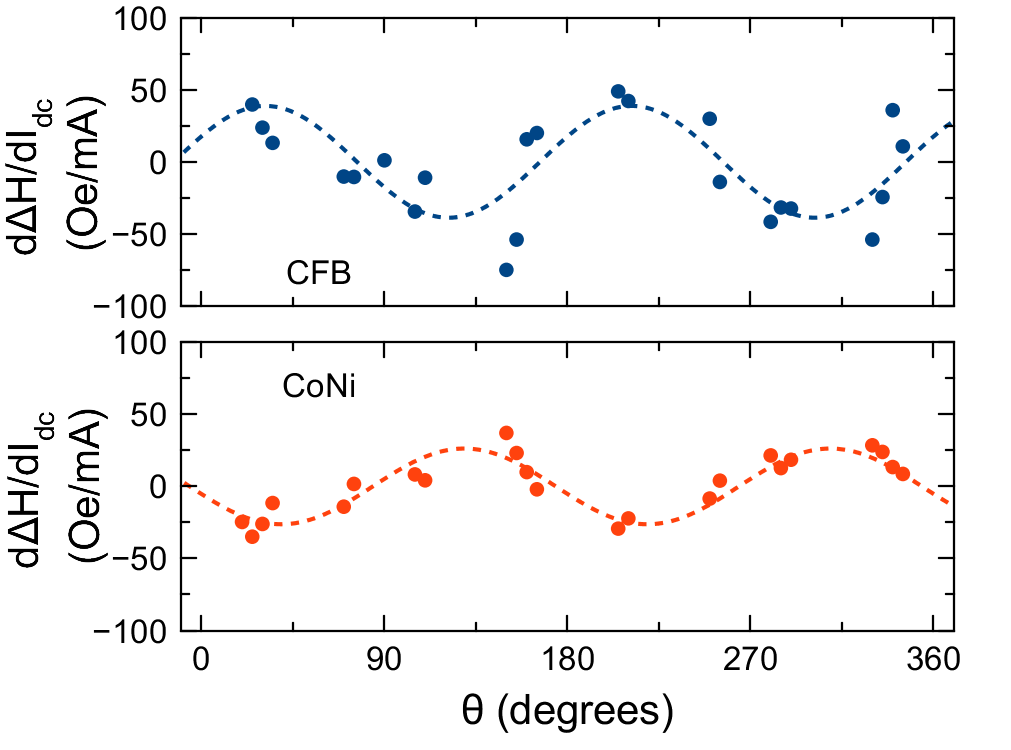}
\caption{
Angular dependence of $d \Delta H/dI_{\mathrm{dc}}$ for the CFB (top) and CoNi (bottom) layers with a fit to the expected angular dependence (dotted line). 
}
\label{fig3}
\end{figure}

To  determine the angular dependence of the  torque, we measure the slope of linewidth vs bias $d \Delta H/dI_{\mathrm{dc}}$ at multiple angles in the $xz$ plane. We exclude angles where the peaks overlap, since fitting overlapping curves   introduces additional error and effects such as dynamic exchange coupling that can modify resonance linewidths \cite{Heinrich2003}.  Figure\,\ref{fig3}  shows the measured slope $d \Delta H/dI_{\mathrm{dc}}$ for both layers. The dotted fit  follows  the expected $\cos(\theta)\sin(\theta)$ angular dependence for PHE driven torques. Other sources of spin current such as spin Hall and  Rashba  effects are known to produce torques as well. However, for the geometry used here,  their spin polarization direction is along $\uvec{y}$ and would result in the $d \Delta H/dI_{\mathrm{dc}}$ angular symmetry   of $(\uvec{m} \cdot \uvec{y})$. This  is inconsistent with our observation.  Spin currents produced by AHE have a polarization following $\uvec{m}$ as well, however the flow direction follows $\uvec{m} \times \uvec{x}$. When magnetization lies in the $xz$ plane as studied here, there is no flow of spin current in the $\uvec{z}$  direction towards CFB \cite{Taniguchi2015}.


The angular dependence of the observed torques is consistent with the absorption of angular momentum in the CFB layer from a spin current produced in the CoNi layer.  In Ref. \cite{Safranski2018}, a similar torque was observed in a single ferromagnet paired with a spin sink. When comparing the torque on the CoNi layer in this work to Ref. \cite{Safranski2018}, the sign  is consist with a larger spin current transfer from CoNi in the direction of the Au layer. While there is a Pt layer on the other interface, its resistivity is high due to the thin nature of the layer \cite{Nguyen2016}. When considering the CFB layer, the symmetry of the torque matches what  would be expected if the layer was receiving a transfer of angular momentum from the CoNi layer, in that its linewidth decreases as the CoNi linewidth increases. While CFB is known to have AMR, the CoNi system has a significantly higher AMR \cite{McGuire1975} than  CFB \cite{Seemann2011} (See Supplementary Note 1). Further, the resistiviity of CFB is higher, resulting is less charge current passing through this layer. As such,  the CoNi layer is expected  to be the main source of planar Hall driven torques. 

We next estimate the strength of the observed effect. Here we aim to determine the relative efficiency described by a dimensionless coefficient $\eta_\mathrm{{FM}}$, similar to a spin Hall angle. We define $\eta_\mathrm{{FM}}$ to represent the conversion efficiency from charge-current to the respective damping-like torque on the CoNi and CFB FM layers. Assuming a collinear geometry and to the leading order, we would expect  the spin current to alter the resonance linewidth linearly for each layer   with a slope:
\begin{eqnarray}
\dfrac{d \Delta H}{dI_{dc}} = \dfrac{\hbar}{2e} \dfrac{\eta_{FM}}{M_{s}t_{FM}} \dfrac{\cos(\theta)\sin(\theta)}{w t_{tot}}
\end{eqnarray}
where $M_\mathrm{s}$ is the saturation magnetization, $t_\mathrm{FM}$ is the particular layer FM thickness, $t_\mathrm{tot}$ is the total stack thickness, and $w$ the bridge width. Taking $M_\mathrm{s}$ to be 800\,emu/cm$^3$ for both layers and using the amplitude of $d \Delta H/dI_{\mathrm{dc}}$ {\it vs} $\theta$ from the fits in Fig.\,\ref{fig3}, we calculate the efficiency of the torque on the CFB layer $\eta_\mathrm{CFB}$ to be 0.05 and for the CoNi layer $\eta_\mathrm{CoNi}$=0.09. Comparing to the data in Ref \cite{Safranski2018},  the effective $\eta$ of the spin Hall in a Pt/CoNi system is 0.07 and for the planar Hall based torques 0.03. This comparison shows that the spin current generation measured here is on the same order as the spin Hall effect in Pt.  


In conclusion, we have observed the action of spin current generation in a CoNi/Au/CFB system in the linewidths of the two FM layers. We find the angular dependence of the spin torque magnitude  to be consistent with the planar Hall effect associated with the CoNi layer. The overall charge to spin conversion efficiency in this structure is on par with the spin Hall effect in Pt. However, unlike the spin Hall effect, here a partially out of plane polarization is achieved. We believe PHE driven torques allow for deterministic switching of perpendicular magnetic layers without the need for symmetry breaking. Such controllable spin polarization will also allow studies of new spin transport configurations.

\section*{Acknowledgment}

Work done with the MRAM group at IBM T. J. Watson Research Center in Yorktown Heights, New York. ADK acknowledges support from the National Science Foundation under Grant No. DMR-1610416. Research at NYU was supported partially by the MRSEC Program of the National Science Foundation under Award Number DMR-1420073. This work was performed in part at the Advanced Science Research Center Nano-Fabrication Facility of the Graduate Center at the City University of New York.

\bibliography{main}

\clearpage
\widetext

\begin{center}
\textbf{\large Supplementary Material for \\
Planar Hall driven torque in a FM/NM/FM system}
\bigskip
\end{center}
\onecolumngrid
\setcounter{equation}{0}
\setcounter{figure}{0}
\setcounter{table}{0}
\setcounter{page}{1}
\renewcommand{\theequation}{S\arabic{equation}}
\renewcommand{\thefigure}{S\arabic{figure}}

\renewcommand{\figurename}{Supplementary Figure}
\renewcommand\refname{Supplementary References}
\def\bibsection{\section*{\refname}} 

\section*{Supplementary Note 1: Magnetoresistance Measurements}

\begin{table*}[h]
	\centering
  	\caption{Summary of film level magnetoresistance determined by 4 point probe measurements. Numbers in parenthesis are layer thickness in nm.  }\label{AMR}
  	\begin{tabular}{lc}
	\hline
 		Sample 																& MR  $(\%)$  \\
        \hline
 		CFB(10.0)/MgO 														 	& 0.04 			\\
 		Ta(3.0)/[Co(0.65)/Ni(0.98)]$_2$/Co(0.65)/Ta(3.0)												 	& 0.4 \\
        Ta(3.0)/Au(3.0)/CFB(1.5)/Ta(3.0) 												 	& 0.009 \\
        Ta(3.0)/Pt(3.0)/Au(3.0)/CFB(1.5)/Ta(3.0) 												 	& 0.008 \\
 		Ta(3.0)/Pt(3.0)/[Co(0.65)/Ni(0.98)]$_2$/Co(0.65)/Au(3.0)													 	& 0.5 \\
        \hline
  \end{tabular}
\end{table*}


The planar Hall effect is strongly related to the anisotropic magnetoresistance (AMR) in metallic ferromagnets\cite{Taniguchi2015,Safranski2018,Seemann2011}. From literature, the AMR in the CoNi system\cite{McGuire1975} should be much larger than that found in CFB\cite{Seemann2011}. In the following section, we measure the strength of the magnetoresistance in the material stacks used in this study. Using four point resistance measurements on thin films, we define the strength to be  $MR = 100(1- \rho_{\mathrm{x}}/ \rho_{\mathrm{z}})$, where $\rho_{\mathrm{x}}$ is the resistivity with a 7\,kOe magnetic field applied along the current path, and $\rho_{\mathrm{z}}$ measured with field applied perpendicular to the sample plane. The MR we report here is related to AMR, but is diluted by the resitivity of the shunting layers. We neglect the resistivity in the $y$ axis, as spin Hall magnetoresistance contributes to the measured  magnetoresistance\cite{Nakayama2013} and is known to be on the order of AMR in metallic structures\cite{Kim2016}.

Supplementary Table\,\ref{AMR} shows the MR strength for various multilayer stacks. The first two entries are a thick CFB layer and the CoNi stack used in the main text without the additional Au and Pt layers. In these measurements, we see that the AMR in CoNi is roughly an order of magnitude larger. However, thin film resitivities and properties are very sensitive to their thickness and the seed layer. We next introduce additional layers underneath the CFB layer with thicknesses identical to the main text. Insertion of a Ta/Au layer reduces the total observed MR in CFB to 0.009$\%$, likely due to shunting of current through the Au layer. The addition of a Pt layer to this stack further reduces the observed MR. Next we measure the CoNi layer with the same Pt and Au shunting layers and observe a much larger MR than with CFB. Further, the MR of CoNi with the shunting layers is larger than CoNi grown on Ta. This indicates that the Pt seed encourages better growth of the CoNi layers, and that the actual AMR is in reality higher than 0.5$\%$. From these MR values, we conclude that the AMR in CFB is negligibly small and the AMR in CoNi is responsible for observed torques. 

\end{document}